\documentclass[%
 aip,
 amsmath,amssymb,
 reprint,%
]{revtex4-1}

\usepackage{graphicx}
\usepackage{dcolumn}
\usepackage{dirtytalk}
\usepackage{bm}

\usepackage[utf8]{inputenc}
\usepackage[T1]{fontenc}
\usepackage{mathptmx}
\usepackage{amsmath}
\usepackage[monochrome]{color}

\setcitestyle{super}

\newcommand*{\citen}[1]{%
  \begingroup
    \romannumeral-`\x 
    \setcitestyle{numbers}%
    \cite{#1}%
  \endgroup   
}

\newenvironment{Notes}
{\begin{quote}\small\tt Note from Ioan: \ }
{\end{quote}}

\newenvironment{Notes_2}
{\begin{quote}\small\tt Note to Ioan: \ }
{\end{quote}}

\newcommand{\bno}{\begin{Notes}}
\newcommand{\eno}{\end{Notes}\noindent}

\begin{document}

\preprint{AIP/123-QED}

\title[]{Weighted Ensemble Milestoning  (WEM): A Combined Approach for Rare Event Simulations}

\author{Dhiman Ray}
\author{Ioan Andricioaei}%
 \email{andricio@uci.edu.}
\affiliation{ 
Department of Chemistry, University of California Irvine, California 92697, USA
}%



\begin{abstract}
 To directly simulate rare events using atomistic molecular dynamics is a significant challenge in computational biophysics. Well-established enhanced-sampling techniques
 do exist to obtain the thermodynamic functions for such systems. But developing methods for obtaining the kinetics of long timescale processes from simulation at atomic detail is comparatively less developed an area. Milestoning and the weighted ensemble (WE) method are two different stratification strategies; both have shown promise for computing long timescales of complex biomolecular processes. Nevertheless, both require a significant investment of computational resources. We have combined WE and milestoning to calculate observables in orders of magnitude less CPU and wall-clock time. Our weighted ensemble milestoning method (WEM) uses WE simulation to converge the transition probability and first passage times between milestones, followed by the utilization of the theoretical framework of milestoning to extract thermodynamic and kinetic properties of the entire process. We tested our method for a simple one-dimensional double well potential, an eleven-dimensional potential energy surface with energy barrier, and on the biomolecular model system alanine dipeptide. We were able to recover the free energy profiles, time correlation functions, and mean first passage times for barrier crossing events at a significantly small computational cost. WEM promises to extend the applicability of molecular dynamics simulation to slow dynamics of large systems which are well beyond the scope of present day brute-force computations.
\end{abstract}

\maketitle

\section{Introduction}
\noindent
Computing both thermodynamics and kinetics for complex processes from molecular dynamics (MD) simulation is of significant interest, but can be a substantial challenge\cite{Klippenstein2014}. For example, biologically relevant molecular processes take place over a wide range of timescales spanning from picoseconds to seconds: protein side chain motion (ps-ns), relative motion of different protein domains (ns-$\mu$s), protein folding, ligand binding and allosteric transitions ($\mu$s-s) \cite{Zwier2010,Votapka2017}. 


In the proper aqueous environment and using femtosecond-step integration\cite{Hopkins2015} it is difficult to "see" events that occur on the $\mu$s-to-second timescale with currently available computational resources. Specialized computing hardware, like the Anton supercomputer \cite{Shaw2008}, can perform $\mu$s-to-ms simulations of systems composed of hundreds of thousand of atoms. Distributed computing over graphical processing units (GPUs) and GPU-grid based hardware has also enabled $\mu$s timescale simulation \cite{Buch2010, Buch2011}. The ms-to-second time scale remains a challenge for regular molecular dynamics simulations. 

Although enhanced sampling methods such as umbrella sampling (US) \cite{Torrie1977}, parallel tempering/replica exchange  \cite{Swendsen1987NonuniversalSimulations,Sugita1999Replica-exchangeFolding}, metadynamics (MtD) \cite{Laio2002}, adaptive biasing force (ABF) \cite{Darve2008} and others (reviewed in Ref. \citen{Armacost2020NovelApplications}, for example) can provide the free energy landscape along pre-defined reaction coordinate, the kinetic properties often elude accurate estimation from MD simulation, because enhanced sampling perturbs the real dynamics of the system. One avenue to estimate kinetics is via rate theories based on the energy landscape barriers estimated at equilibrium.  Examples are temperature accelerated MD \cite{Srensen2000Temperature-acceleratedEvents,Zamora2016TheApproach} and hyperdynamics \cite{Voter1997Hyperdynamics:Events} proposed by Voter, as well as conformational flooding by Grubmuller and co-workers\cite{Grubmuller1995PredictingFlooding} and bias-exchange metadynamics by Piana and coworkers. \cite{Marinelli2009ASimulations} Another example is from Tiwary and Parinello, who have recently devised a scheme for calculating rate constants from infrequent metadynamics simulations, using an Arrhenius rate equation based approach \cite{Tiwary2013}. This method has been proved to be successful for $\mu$s scale ligand unbinding and folding simulations of small proteins \cite{Wang2018}. Alternatively, Doshi and Hamelberg \cite{Doshi2011ExtractingTheory}, and Frank and Andricioaei have developed a method of estimating rate constants of slow biomolecular processes using Kramers rate theory \cite{Kramers1940BrownianReactions} from potential scaled simulations \cite{Frank2016}. This method has been used for determining first passage times of drug unbinding from protein from $\mu$s scale simulation \cite{Deb2019AcceleratingSimulations}.  

The fundamental challenge for obtaining converged kinetics of the conformational change between states for a slow process is sampling many successful transitions from the initial state to the final state(s) \cite{Grazioli2018a}; this requires significantly longer MD simulation time than the mean first passage time of the rarest event. As most processes of physiological interest happen the beyond $\mu$s timescale, simulating many such events goes well beyond the scope of currently available computational resources.

This is because one single long MD trajectory is prone to remain trapped in low energy minima and therefore avoids climbing over high energy regions \cite{Bagchi2018StatisticalScience}. Consequently, it ends up oversampling local minima close to the starting configuration, but fails to cross the rate-limiting free energy barriers to reach the target state for purely statistical reasons \cite{Laio2002}. To improve statistics, one can run multiple trajectories with different initial conditions, but this leads to a significant increase in computational cost.

Path sampling, i.e., strategies based on the statistical mechanics of the paths followed during the transitions of interest, \cite{Dellago1998,Bolhuis2002} have proved to be a convenient framework for estimating kinetics of rare events from MD simulations \cite{Chong2017}.  Within this framework, a useful general approach is to stratify the sampling of trajectories (see Ref.\cite{Dinner2018TrajectoryDynamics} for an insightful description of stratification). Instead of a continuous, long trajectory, it is computationally more expedient to generate many short trajectories inside strata of configuration space and then piece them together with proper path statistics. In this vein, a variety of methods like transition interface sampling (TIS) \cite{VanErp2003}, forward flux sampling (FFS) \cite{Allen2006,Allen2009}, weighted ensemble (WE) \cite{Huber1996,Bhatt2010,Zhang2010} and milestoning \cite{Faradjian2004,West2007,Bello-Rivas2015} have been designed for equilibrium kinetics or specifically for non-equilibrium steady states \cite{Dickson2009SeparatingSampling,Warmflash2007UmbrellaProcesses,Vanden-Eijnden2009ExactTilting}. 

These methods have in common the stratification strategy: they discretize the reaction coordinate into bins (or intermediate states, or strata). In the WE and FFS methods, MD trajectories are stopped and spawned if they cross the bin boundaries, increasing the number of trajectories in relatively under-sampled regions of the configuration space. In TIS and milestoning, trajectories are initiated from intermediate values of the reaction coordinate and the probabilities that they reach the initial, final or other intermediate interfaces are monitored. These intra-strata trajectories can also be accelerated by applying biasing force that push away from the starting milestone towards the neighboring ones\cite{Grazioli2018a}; proper statistics is recovered by re-weighting based on stochastic action path integrals to get the appropriate kinetic information \cite{Zuckerman2000EfficientDimension,Nummela2007,Xing2006,Perilla2011ComputingSampling}.



Utilizing multiple short trajectories as input data, Markov state modeling (MSM) can also be thought of as a way to stratify configuration space. MSM decomposes a large amount of trajectory data into meta-stable states or clusters, depending on structural or kinetic criteria, and subsequently builds a Markov chain between the states \cite{Pande2010,Bowman2009,Noe2008,Bowman2009a}.  The primary advantage of MSM is that it does not require a predefined reaction coordinate. Techniques like time-lagged independent component analysis (TICA) can be used to identify the slowly varying degrees of freedom \cite{Perez-Hernandez2013,Schwantes2013}. The thermodynamics and kinetics along them can later be captured from the eigenvalues and eigenvectors of the Markovian transition probability matrix.

Recently, milestoning and weighted ensemble methods have gained popularity in computational biophysics because of their open source implementation in commonly used molecular dynamics packages. 
For example, the Weighted Ensemble Simulation Toolkit with Parallelization and Analysis (WESTPA) performs WE simulation in conjunction with molecular dynamics and Brownian dynamics \cite{Zwier2015}. It has found its applications in studying the free energy and kinetics of a plethora of interesting biophysical processes including protein folding \cite{Abdul-Wahid2014,Adhikari2019ComputationalTimes}, formation of host-guest complexes \cite{Zwier2011}, protein ligand binding \cite{Zwier2016} \textcolor{black}{and unbinding \cite{Dickson2016}}, ion permeation through protein channels \cite{Adelman2015}, viral capsid assembly \cite{Spiriti2015} etc. \textcolor{black}{Using $\mu$s-long WE simulation, Adhikari \textit{et al.} have calculated the rates of millisecond to second timescale unfolding of small proteins (NTL9 and Protein G) in implicit solvent at room temperature \cite{Adhikari2019ComputationalTimes}. Their calculations were within an order of magnitude from the experimental numbers. Previous calculation of the folding rate of the same proteins could only be carried out at a higher temperature because room temperature experimental timescales were not accessible in conventional MD simulation \cite{Lindorff-Larsen2011HowFold,Voelz2010MolecularNTL91-39}. In the same vein, Saglam and Chong have reported an explicit solvent simulation of large scale protein-protein binding of multi-microsecond timescale using WE methodology \cite{Saglam2019Protein-proteinSimulations}.  They could accurately reproduce the experimental rate constants spending only $<$1\% of the simulation time required to construct a converged MSM \cite{Plattner2017CompleteModelling}, a state of the art technique to calculate biomolecular kinetics. 
Lotz and Dickson \cite{Lotz2018,Dickson2017MultipleWExplore} have calculated the kinetics of an \emph{11-minute} drug unbinding process using their WExplore implementation \cite{Dickson2014WExplore:Algorithm} of the WE scheme, although the ligand residence time they obtained (42 seconds) was more than one order of magnitude off from the experimental timescale. All these results collectively bolster the growing  utility of WE based methods in MD simulation of rare events, and, at the same time, point towards the need to push methodological boundaries farther in timescale.} 

At the same time, Elber and coworkers 
have recently shown the rigorous derivation of expressions for the free energy landscape and mean first passage time from milestoning, a method they proposed in the last decade. They also proved that milestoning can be considered statistically exact at the infinite sampling limit \citep{Bello-Rivas2015}. Different variants of milestoning have incorporated innovative strategies for constructing milestoning boundaries (e.g., Voronoi tessellations) to effectively sample the reaction coordinate on the way to the product state. Examples include the directional milestoning of Majek et al. \cite{Majek2010} and the Markovian milestoning by Vanden-Eijnden et al. \cite{Vanden-Eijnden2009}. Taken collectively, the milestoning methods have been successfully applied to problems like allosteric transitions \cite{Elber2007}, membrane permeation by small molecules \cite{Votapka2016,Fathizadeh2019} and other biological membrane systems \cite{Cardenas2016,Vanden-Eijnden2009}. Votapka et al. have simulated the binding of a ligand to a protein using multi-scale simulations involving Brownian dynamics and molecular dynamics in conjunction with milestoning \cite{Votapka2015}. They have implemented their technique in a open source package SEEKR which can run alongside the MD simulation package NAMD \cite{Phillips2005} and the BD package BrownDye \cite{Huber2010Browndye:Dynamics,Votapka2017}. In our own work, we have developed an enhanced milestoning protocol by applying biasing force to obtain quick convergence for milestone-to-milestone trajectories \cite{Grazioli2018a}, and we have also devised a strategy to extract time correlation functions from milestoning simulations using stochastic path integrals \cite{Grazioli2018}, a strategy which can be used in principle for any trajectory stratification technique.



Currently, the WE method generally aims at converging to a non-equilibrium steady-state, from which one can estimate the rate constant using the probability flux into the product state; one can also generate from WE possible reactant to product transition pathways. \textcolor{red}{In this protocol, the trajectories reaching the final state are to be regenerated from the starting point to maintain a steady state of probability flow \cite{Bhatt2010} (except for a recent study which shows that transient flux information can also be used to obtain reasonable kinetics \cite{Adhikari2019ComputationalTimes}). Using this trajectory regeneration approach, the binding of MDM2 protein with an intrinsically disordered p53 peptide has been studied by Zwier et al. \cite{Zwier2016} and  $k_{on}$ has been estimated with an accuracy that was within one order of magnitude from the experimental value. Recently, Saglam and Chong have calculated the association rate constant between barnase and barstar proteins within the error of the corresponding experiment \cite{Saglam2019Protein-proteinSimulations}. Dickson and Lotz could also obtain converged dissociation kinetics and the exit point distribution of small molecule ligands from a protein target using their steady state WE scheme (WExplore) \cite{Dickson2014WExplore:Algorithm} within a fraction of computational cost for conventional simulation.  \cite{Dickson2016}.}

However, when obtaining the free energy profile and kinetics is the sole objective, an equilibrium version of WE can be used without recycling the trajectories from the sink state \cite{Zhang2010}. \textcolor{red}{This ``equilibrium'' version of WE has been developed and used by Suarez et al. \cite{Suarez2014SimultaneousTrajectories} to calculate the free energy surface and timescales of conformational transitions using a history dependent Markov scheme. As an additional advantage, the definition of the target state can be modified after the simulation, removing any artificial bias caused by arbitrarily chosen states. The rate constants can then be recovered by dividing the equilibrium WE trajectories into two sets (depending on their origin from initial or final states) and computing the steady state flux for each set or direction \cite{Suarez2014SimultaneousTrajectories}.} 

This paper primarily focuses on the latter approach, and tries to improve its convergence by utilizing the additional parallelization provided by milestoning. Essentially, the method we put forth combines the efficiency of WE with the milestoning framework for using discontinuous trajectories to estimate rate constants and free energy profiles. 

\textcolor{black}{A limitation of WE simulation is that it can generate correlated trajectories, as a part of their history is common. While trajectories which are not independent are prone to give results with large variance, this correlation does not however lead to bias \cite{Zuckerman2017}. Also, while in principle the WE formalism can provide correlation functions, to the best of our knowledge, there has been no study on time-correlation functions obtained from WE-based methods.}


Milestoning, on the other hand, has precise ways of calculating free energy, kinetics and time correlation functions from the milestone-to-milestone transition probabilities and lifetimes. However, milestones have to be placed sufficiently far from each other to ensure that trajectories starting from a given milestone are independent from their history of previously visited milestones \cite{Majek2010}. This requires longer simulation timescales compared to closely spaced interfaces, and effectively incurs a large computational cost when summed over all milestones. The wind-assisted re-weighted milestoning (WARM) attempts to reduce some of this cost by accelerating the trajectories towards the adjacent milestone by applying a biasing force   \cite{Grazioli2018a}. 

In the current work, we propose a novel path stratification algorithm combining weighted ensemble and milestoning to produce a computationally more efficacious technique than either of the individual techniques. The essence of the method is to perform \emph{weighted ensemble simulations in-between milestones} for fast convergence of the transition probability within each segment between the milestones. The converged  transition probability matrix is then utilized to calculate the free energy, kinetics and time correlation function from the framework of milestoning theory described in Ref. \citen{Bello-Rivas2015}.

The rest of the paper is organized in the following manner. In Section \ref{sec:theory} we review the theoretical framework of milestoning and weighted ensemble in brief, and proceed to describe our combined weighted-ensemble milestoning (WEM) procedure. In Section \ref{sec:app} we show the results of our method on a 1-dimensional double well potential, a coupled (10+1)-dimensional potential with enthalpic barrier and a conformational transition in alanine dipeptide. We compare the results of WEM with WE, milestoning and regular MD simulation wherever applicable. We conclude with a discussion for broader applications of the weighted-ensemble milestoning method. 

\section{Theory}
\label{sec:theory}
\subsection{Milestoning}
\label{sec:milestoning}
The milestoning method was first proposed by Elber and coworkers\cite{Faradjian2004,West2007}, who have recently revisited its approximations and proposed the `exact milestoning' formalism, with rigorous derivations of free energy and kinetics\cite{Bello-Rivas2015}. Here, we include a brief description of the key equations from milestoning which are directly relevant to the our work. The reader is referred to Refs. \citen{Votapka2017} and \citen{Bello-Rivas2015} for the details of the derivations.

Milestones are non-interacting hypersurfaces, ideally orthogonal to the reaction coordinate, that stratify a given phase space \cite{West2007}. $M$ milestones (including the initial and final state) divide the configurational space into $M-1$ domains. The primary goal of milestoning is to estimate the flux of probability $q_{i}(t)$ through the milestone $i$ for $i \in [1,M] $. To accomplish this, multiple trajectories are initiated from each milestone. The trajectories are stopped when they reached either of the adjacent milestones. A transition kernel $\mathbf{K}$ is constructed from the probabilities of transition between adjacent milestones,
\begin{equation}
\begin{split}
K_{ij} & = \frac{n_{i\rightarrow j}}{N}; \qquad j = i \pm 1 \\
& = 0; \qquad \text{otherwise}
\end{split}
\label{eqn:kij}
\end{equation}
where $n_{i\rightarrow j}$ is the number of trajectories initiated at milestones $i$ and ending at milestone $j$, and $N$ is the total number of trajectories started from milestone $i$. Also, a lifetime vector $\mathbf{\overline{T}}$ is obtained; it contains the average lifetime of each milestone. The average lifetime of a milestone $i$ is defined as the average time spent by the trajectories initiated from $i$ before they reach either milestone $i-1$ or $i+1$. Therefore, the elements of $\mathbf{\overline{T}}$ are given by 
\begin{equation}
\overline{T}_i = \frac{\sum_{l=1}^N t_l}{N},
\end{equation}
where $t_l$ is the time spent by the $l$'th trajectory before hitting any of the adjacent milestones. To compute the free energy profile along the reaction coordinate the stationary flux vector $\mathbf{q}_{\text{stat}}$ has to be computed. The stationary flux vector is the \textcolor{black}{left} eigenvector of the transition kernel $\mathbf{K}$ with eigenvalue 1.
\begin{equation}
\mathbf{q}_{\text{stat}}^T \mathbf{K} = \mathbf{q}_{\text{stat}}^T
\end{equation}
\textcolor{black}{Although it is expected that the principal eigenvector of a transition kernel would provide the stationary probabilities at the milestones, one should note that, in steady state, the flux through a given milestone is proportional to the density of trajectory \textit{hitting points} on that milestone. The elements of $\mathbf{q}_{\text{stat}}$ vector are these stationary probability densities of hitting points on different \textit{hyper-surfaces} which is not contradictory to traditional master equation framework.} The equilibrium probability distribution $\mathbf{P_{\text{eq}}}$ at the milestones is obtained from the stationary flux vector by element-wise multiplication of $\mathbf{q}_{\text{stat}}$ with the lifetime vector $\mathbf{\overline{T}}$.
\begin{equation}
P_{\text{eq},i} = q_{\text{stat},i} \overline{T}_i
\label{eqn:peq}
\end{equation}
\textcolor{black}{By the probability at milestone $i$ we refer to the probability of a trajectory that has last crossed milestone $i$ and did not cross any other milestone afterwards \cite{Bello-Rivas2015}. (It is different from trajectory \textit{hitting point} density, we mentioned before.) } From the equilibrium probability distribution the free energy $\Delta G_i$ at milestone $i$ can be calculated as
\begin{equation}
\Delta G_i = -k_BT \ln \bigg(\frac{P_{\text{eq},i}}{P_{\text{eq},0}} \bigg)
\label{eqn:free-energy}
\end{equation}
where $P_{\text{eq},0}$ is the probability corresponding to a reference free energy.

To calculate the mean first passage time (MFPT) $\tau$, an absorbing boundary condition has to be set at the last milestone, the target state \cite{Bello-Rivas2015}. The milestone at the endpoint works as a sink and the transition probability from that milestone to any other milestone is set to zero. So we define a new square matrix $\mathbf{\tilde{K}}$ with the property that
\begin{equation}
\begin{split}
\tilde{K}_{ij} & = 0;\qquad \text{if}\; i \geq m, j = i \pm 1 \\
& = K_{ij};\qquad \text{otherwise}
\end{split}
\label{eqn:ktilde}
\end{equation}
where milestone $m$ is referring to the final state of the problem in hand. It is possible that there are milestones after $m$ and $\tilde{K}_{ij}$ will be zero for them as well. Now, the MFPT is calculated from
\begin{equation}
\tau = \mathbf{p_0}(\mathbf{I}-\mathbf{\tilde{K}})^{-1}\mathbf{\overline{T}}
\label{eqn:mfpt}
\end{equation}
The derivation of the Eq. (\ref{eqn:mfpt}) is presented in Ref. \citen{Bello-Rivas2015}. The $\mathbf{p_0}$ is the probability distribution at each milestone at the beginning. For example, if we want to study the transition from milestone 2 to milestone 5 for a 5 milestone system we want the initial probability to be at milestone 2 is 1 and zero elsewhere, so $\mathbf{p_0} = (0,1,0,0,0)$. 

Previously, we have shown that time correlation functions can also be calculated by using the transition kernel and the first passage time distribution between milestones \cite{Grazioli2018}. For that, a move from milestone $i$ to $j$ is proposed with probability $p_{i\rightarrow j}$, where
\begin{equation}
    p_{i\rightarrow j} = \frac{K_{ij}}{\sum_{j \in \{ i-1,i+1\}} K_{ij}}
\end{equation}
The time taken for the move is a random number sampled from the first passage time distribution between those given milestones. This process is repeated many times to generate a trajectory in the milestone space, 
\begin{equation}
\mathbf{X} = \lbrace (t_n,x_i) \mid n \in [1,N]; i \in [1,M] \rbrace,
\end{equation}
where $M$ is the number of milestones, $N$ is the total number of steps $n$, and $x_i$ is the value of the coordinate corresponding to the milestone $i$. This method is akin to a kinetic Monte Carlo scheme \cite{Voter2007INTRODUCTIONMETHOD}. The time $t_n$ is defined as 
\begin{equation}
t_n = \sum_{k=1}^N t^s_k
\end{equation}
where $t^s_k$ is the sampled first passage time for $k$th move. This long trajectory $\mathbf{X}$ can be extended much beyond the timescales achievable from conventional MD simulation. There are two ways to calculate time correlation functions over the dynamics described by $\mathbf{X}$. One is to construct a time dependent conditional probability distribution function $P_i (t\mid x_0)$ by interpolating the sparse time trajectory $\mathbf{X}$ for all milestones. Here $i \in [1,M]$ and $x_0$ is the initial position. From this distribution, a time correlation function can be estimated by evaluating the following expression \cite{Grazioli2018}
\begin{equation}
C(t) = \sum_{i=1}^M \bigg(A(x_i)P_i(\infty) \sum_{s=1}^M A(x_s)P_s(t \mid x_0)\bigg)
\end{equation}
where $A$ is the observable, $x_i$ is the position at milestone $i$ and $\mathbf{P}(\infty)$ is the all-milestone stationary probability distribution vector obtained from Eq. (\ref{eqn:peq}).

The other option is to interpolate the long trajectory $\mathbf{X}$ to get positions and times in between milestones to produce another trajectory of the same length but with much higher resolution in time and space. From this trajectory the time correlation can be calculated by conventional time averaging 
\begin{equation}
C(t) = \frac{1}{T}\int_0^T A(x(t^{\prime}))A(x(t^{\prime}+t))dt^{\prime}
\label{eqn:time-corr}
\end{equation}
In our previous work \citep{Grazioli2018}, we have shown that the latter method performs better for effectively low dimensional systems. When applied to alanine dipeptide (two important degrees of freedom), it yielded bond re-orientation time-correlation function (using the Lipari-Szabo formalism for NMR order parameters \cite{Lipari1982Model-freeValidity}) in good agreement with the exact results. However, our proposed formalism involved one long brute-force Langevin trajectory to begin with, which was discretized into the milestone space. We did not implement this technique for short milestone to milestone trajectories, we add this implementation herein.

\subsection{Weighted ensemble (WE)}
Weighted ensemble simulation is a statistically exact path stratification strategy introduced by Huber and Kim \cite{Huber1996}, and later studied in detail by Zuckerman and coworkers \cite{Bhatt2010,Zhang2010,Zuckerman2017}. The details of the weighted ensemble simulation are presented in Ref. \citen{Bhatt2010}. A brief discussion regarding details relevant to our work is given below.

In WE simulation the configuration space is divided into multiple (say $M$) bins. A certain number of trajectories (say $N$) are started from the initial state. Each trajectory has weight $1/N$ in the first iteration. The position is monitored at a given time interval $\delta t$. After each time interval any trajectory reaching a new bin is stopped and some new trajectories are generated from its endpoint. The weight of the old (parent) trajectory is equally distributed amount the newly created (daughter) trajectories. This process is continued so that every occupied bin will contain exact $N$ trajectories. If the number increase beyond $N$, the excess ones are \textcolor{black}{stopped} and their weights are added to the surviving trajectories. Thus the total probability remains conserved. The philosophy of the WE simulation is that 
the number of trajectories gets increased and the weights get reduced as we go further from the starting state. Some of this trajectories with very low weight might eventually reach the target state. From these weights it is possible to estimate the rate and mean first passage time. For example if a trajectory with weight $10^{-6}$ reaches the target state at certain time $t$ then it is possible to suggest that in time duration $t$, approximately 1 out of $10^6$ trajectories reaches that target state without propagating all of them during the entire course of time. To maintain a steady state of probability flow, the trajectories reaching the target state are re-initiated from the starting state. The rate constant $k$ and mean first passage time (MFPT) can be estimated by the ``Hill relation": \cite{Zuckerman2017}
\begin{equation}
k = \frac{1}{\text{MFPT}(A\rightarrow B)} = \text{flux}(SS,A\rightarrow B) = \frac{\sum_i w_i}{t}
\end{equation}
where $SS$ refers to steady state, $A$ and $B$ are the initial and target states respectively, $w_i$ is the weight of the $i$th reactive trajectory and $t$ is the simulation time.

\subsection{Combining weighted ensemble with milestoning (WEM)}
In the WEM method we blend milestoning and WE methods into one so that the convergence of milestone-to-milestone transition probabilities and timescales can be achieved with reduced computational effort. We divide the space between milestones into bins and apply the WE technique of \textcolor{black}{splitting and merging} trajectories in between the milestones. The $i$-th milestone becomes our starting state of the WE simulation and milestones $i-1$ and $i+1$ become two target states. A schematic of our method is shown in Figure \ref{fig:schematic}. 
 
Our WEM strategy introduces the following modifications to the weighted ensemble scheme: 
\begin{figure}
    \centering
    \includegraphics[scale=0.3]{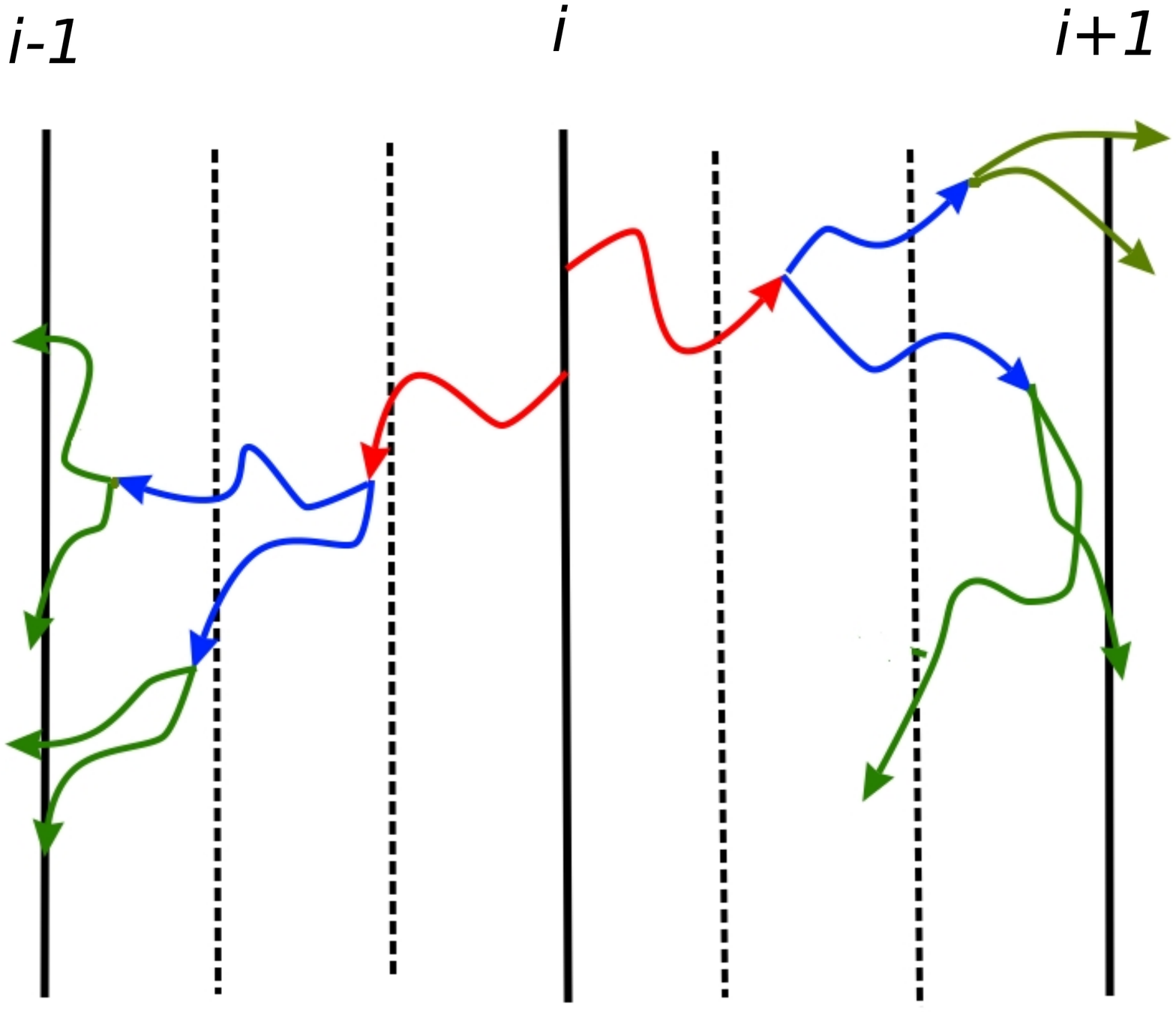}
    \caption{Schematic diagram of the combined, weighted ensemble milestoning (WEM) method. The solid vertical lines are milestones and the dotted lines are WE bin boundaries. The trajectory in red is the starting trajectory which branches out into daughter trajectories upon crossing WE bin boundaries. The blue and green trajectories have weights $1/2$ and $1/4$ of the red one, respectively. In the example of this figure, $K_{i\;i+1} = \frac{3}{8}$ and $K_{i\;i-1} = \frac{1}{2}$. }
    \label{fig:schematic}
\end{figure}
\begin{itemize}
\item The trajectories reaching either of the adjacent milestones are \textcolor{black}{stopped}. 
\item No new trajectories are generated when a trajectory reaches a target milestone, because we do not enforce a (non-equilibrium) steady state condition \textit{per se} in our simulation. This is typical for conventional milestoning.
\item The average lifetime of milestone $i$ is estimated by 
\begin{equation}
\overline{T}_i = \sum_k t_k w_k
\end{equation}
where $w_k$ and $t_k$ are, respectively, the weight and time at which the $k$th trajectory reaches either milestone $i-1$ or $i+1$.
\item The elements of $\mathbf{K}$ (from Equation \ref{eqn:kij}) are given by
\begin{equation}
\begin{split}
K_{ij} & = \sum_{k\in \Gamma(i\rightarrow j)} w_k; \quad j = i \pm 1 \\
& = 0;\quad \text{otherwise}
\end{split}
\label{eqn:kij_wem}
\end{equation}
where $\Gamma(i\rightarrow j)$ is the set of trajectories starting at milestone $i$ and ending at $j$. If all the trajectories reached either of the milestones, then $K_{i\;i+1} + K_{i\;i-1} = 1$ \textcolor{black}{(right stochasticity)}. 


\item The first passage time distribution ($FPTD_{i \rightarrow j} (t)$) is calculated as a function of time by summing over the weights of the trajectories arrived at $j$ from $i$ within the time interval of $t$ and $t+\delta t$. 

\textcolor{black}{\item The right stochasticity criteria can be used as a check for convergence for each milestone. Yet, if the milestone $i$ is placed in a deep free energy minimum, this condition might not be satisfied. So, we have used the first passage time distribution function as a measure of convergence. When the first passage time distribution of transition to both the nearby milestones goes below a tolerance (which we chose to be the probability value $10^{-4}$ for all our calculations) we deem the simulation from that particular milestone converged.}
\end{itemize}

The transition kernel, lifetime vector and first passage time distribution calculated in this manner are used in the milestoning theory as described in Section \ref{sec:milestoning} to elucidate the MFPT $\tau$, stationary probability distribution (which gives the free energy profile), and time correlation functions for the overall process.

\section{Results}
\label{sec:app}
\subsection{One Dimensional Double-Well potential}
\label{sec:1d}
\begin{figure}

\includegraphics[scale=0.5]{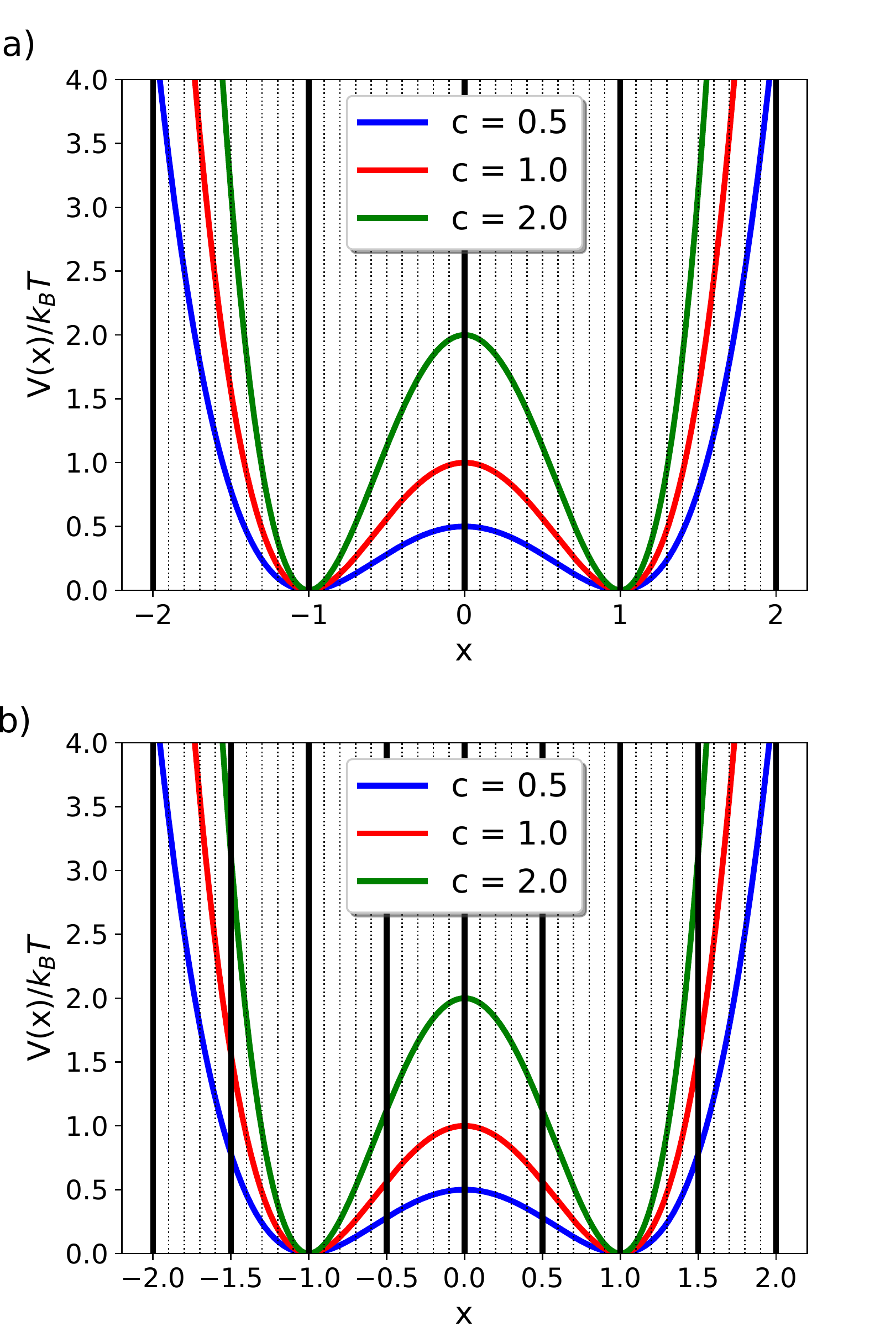}
\caption{The 1D double well systems with different barrier heights and milestone configurations. The positions of the milestones are depicted by solid black lines and the edges of the WE bins are depicted by black dotted lines. (a) 5 milestone system; (b) 9 milestone system}
\label{Potential}
\end{figure}

As proof of principle, we tested our WEM method on a 1D double well potential. The chosen potential is of the form:
\begin{equation}
V(x) = c(1-x^2)^2,
\label{eqn:vx}
\end{equation} 
where $c$ is a parameter which can be varied to control the barrier height (Fig. \ref{Potential}). Such functional forms were used as test cases 
in earlier studies in milestoning \cite{Grazioli2018a,Grazioli2018} and weighted ensemble simulations \cite{Zhang2010}. We have used $k_BT$ as the unit of energy, so the barrier height is $ck_BT$ in Eq. (\ref{eqn:vx}). We have performed WEM simulations for three different barrier heights: 0.5 $k_BT$, 1.0 $k_BT$ and 2.0 $k_BT$. 
Two different milestoning schemes with 5 and 9 milestones have been tested (Fig. \ref{Potential}). The milestones were placed between $x= -2.0$ and $x=2.0$ at $\Delta x=1.0$ interval for the former and $\Delta x = 0.5$ interval for the latter. We studied the transitions between the two minima situated at $x = \pm 1$. Classical trajectories were propagated using over-damped Langevin dynamics \cite{Bhatt2010}:
\begin{equation}
x(t+\Delta t) = x(t) - \frac{\Delta t}{m \gamma} \bigg(\frac{d V}{d x} \bigg)  + \Delta x_R
\label{eqn:odld}
\end{equation}

Time step $\Delta t$ is chosen to be unity. The frictional coefficient $\gamma$ determines the variance of Gaussian white noise $\Delta x_R$ through the fluctuation-dissipation theorem by the following expression \cite{Bhatt2010}
\begin{equation}
\sigma^2 = 2 \frac{k_BT }{m \gamma} \Delta t
\end{equation}
The value of $\gamma$ was chosen to be 2000 which resulted in $\sigma^2=0.001$. Typical values and the propagation algorithm were taken from Ref. \citen{Bhatt2010}. Multiple Langevin trajectories starting from $x=-1.0$ and ending at $x=1.0$ were used to calculate mean first passage time which is compared with that from other milestoning and weighted ensemble simulations. Additionally, a long trajectory of $10^6$ steps has been propagated to calculate the time correlation function for each of the different barrier heights. 

Weighted ensemble (WE) simulations were performed using the WESTPA package \cite{Zwier2015}. The $x$ space has been discretized into bins of width $\delta x = 0.1$. The position of the bins are depicted in Figure \ref{Potential} with dotted lines. WE trajectories were started from $x=-1.0$ and ended at $x = 1.0$. The trajectories were stopped when they reach the target state i.e $x = 1$. $N=10$ trajectories were run in each bin for barrier height $0.5 k_BT$ and 20 for the other two systems. \textcolor{black}{Trajectories were replicated or recombined} at a time interval of $\delta t = 20$ time-steps. Total 2000 iterations of time interval $\delta t$ has been performed. First passage times were calculated by recording the time step at which the trajectories reach the end-point. 

\textcolor{black}{We have chosen the milestones and WE bins arbitrarily and did not put any effort into optimizing their positions. But we did try using different time intervals ($\delta t$) and number of trajectories per bin to obtain converged results with minimum computational cost. The same is true for the other two systems ((10+1) D coupled potential and alanine dipeptide) studied in this paper. } 

In our combined, weighted-ensemble milestoning (WEM) method,
the same $\delta x$ and $\delta t$ values were used as of normal WE simulations. A maximum of 20 trajectories were run for each bin which was sufficient to give converged results for all the double well systems. The results were analyzed using the $w\_ipa$ module of the WESTPA package to elucidate the lifetime of each milestone and the transition probabilities to the nearest milestones. 

Conventional milestoning simulations were also performed to compare with WEM results. The positioning of milestones, number of trajectories, and simulation timescales were kept the same as in the WEM scheme. 

Three different observable quantities were calculated from these simulations: the mean first passage time (MFPT), the stationary distribution along the milestone space, and the position auto-correlation function $C(t) = \langle x(0)x(t) \rangle$. MFPTs are directly related to the rate constant of the transition between the two minima. The stationary distribution can be used to calculate the free energy profile along the chosen coordinate. Finally, the time correlation function is the central quantity in non-equilibrium, characterizing the decay of memory. The mean first passage times for different barrier heights have been computed from milestoning and WEM simulations using Eq. (\ref{eqn:mfpt}). The results are in Table \ref{tab:MFPT}. All error bars are \textcolor{black}{95\% confidence intervals} computed from three sets of independent WEM, WE or milestoning simulations, unless otherwise stated. 

%
The results of the combined weighted ensemble milestoning (WEM) are in mutual agreement with the results from conventional Langevin dynamics simulation, WE simulation, and the conventional milestoning simulation for different barrier heights. However, the WEM simulations require significantly less simulation steps (i.e. total number of force evaluations) to achieve convergence (Table \ref{tab:time}). Although the acceleration compared to traditional WE simulation is not significantly pronounced for this simple 1D system, converged results could be obtained with 2-10 times less simulation steps. 

\textcolor{black}{However, just comparing the total simulation time may not be fair as uncertainties are typically different for results obtained from different methods. So we used a metric which takes into account the spread of the data along with the total simulation time. The algorithm efficiency $\eta$ is defined by
\begin{equation}
    \eta = \left( \frac{A}{\Delta A} \right)^2 N_s^{-1} 
    \label{eqn:efficiency}
\end{equation}
where $A$ is an observable we want to measure and $\Delta A$ is the spread of the measurement. $N_s$ is the number of force evaluations performed to obtain the result. Basically $\eta^{-1}$ denotes the number of force evaluations required to obtain the value of $A$ with an order of magnitude accuracy. \cite{Huber1996}. The values of $\eta^{-1}$ are also shown in Table \ref{tab:time}.  The results indicate that the algorithmic efficiency is poorer for WEM simulation for this simple 1D system. There is not much difference of $\eta^{-1}$ between regular Langevin dynamics and WE simulation. It re-establishes the fact that this 1D system is not a good example for measuring efficiency of any given method. But it shows that we can obtain accurate numbers for the equilibrium probability (explained later) and kinetics from the WEM method.}

It should also be noted that the simulations starting from different milestones are independent and can be run in parallel resulting in significant reduction in wall clock time. We expect this gain to scale significantly better for many-dimensional systems because WEM projects onto important degrees of freedom, hence effectively enhancing motion in the "slow" manifold (see also (10+1)-dimensional Section below).

One should not compare the number of total force evaluations performed for conventional milestoning and WEM simulations because conventional milestoning simulation failed to sample some of the energetically unfavorable transitions, particularly those at the boundaries. We tried increasing the number of walkers and total simulation time by 1-2 orders of magnitude, but we could not see any transition to very high energy milestones. In order to converge statistics for such transitions we may need many orders of magnitude more simulation time or number of walkers; this has been avoided as it is not very relevant for our study.  Because of this under-sampling issue, we did not perform the kinetic Monte-Carlo simulation (see Section \ref{sec:milestoning}) from the conventional milestoning results for calculating time correlation function. \textcolor{black}{We also did not report the algorithmic efficiency of milestoning in Table \ref{tab:time}.} 
\begin{table*}
\caption{\label{tab:MFPT} First passage times in ($\times 10^3$ steps) for different simulation scheme.}
\begin{ruledtabular}
\begin{tabular}{ccccccc}
Barrier height ($k_BT$) &WEM 5 milestone &WEM 9 milestone 
&Regular 5 milestone &Regular 9 milestone
&WE &Regular Langevin\footnote{Error bars for regular Langevin dynamics results are \textcolor{black}{95\% confidence interval of 10000} transition events observed in long Langevin dynamics simulation.}\\
\hline
0.5 &7.5$\pm$1.7 &7.2$\pm$2.0 &6.8$\pm$0.25 &7.3$\pm$0.5 &7.0$\pm$0.25 &6.7$\pm$0.1\\
1.0 &7.0$\pm$1.0 &7.3$\pm$1.0 &7.3$\pm$0.7 &7.8$\pm$1.0 &7.5$\pm$0.25 &7.2$\pm$0.1\\
2.0 &9.6$\pm$0.7 &13.7$\pm$4.2 &9.4$\pm$0.25 &10.3$\pm$3.7 &10.0$\pm$0.25 &10.4$\pm$0.25\\

\end{tabular}
\end{ruledtabular}

\end{table*}

\begin{table*}
\caption{\label{tab:time} Total number of simulation steps (Number of trajectories $\times$ time steps $\times$ number of milestones) required to get converged result. ($\times10^6$ time steps) \textcolor{black}{The inverse of algorithmic efficiency ($\eta^{-1}$) (in the same unit) are included in the parentheses.}}
\begin{ruledtabular}
\begin{tabular}{ccccccc}
Barrier height ($k_BT$) &WEM 5 milestone &WEM 9 milestone &Regular 5 milestone &Regular 9 milestone
&WE &Regular Langevin\\
\hline
0.5 &15.30 \textcolor{black}{(0.786)} &4.86 \textcolor{black}{(0.375)} &16.72 &23.77 &30.66 \textcolor{black}{(0.039)} &\textcolor{black}{66.2 (0.015)}\\
1.0 &18.20 \textcolor{black}{(0.371)} &4.82 \textcolor{black}{(0.090)} &16.72 &23.77 &56.28 \textcolor{black}{(0.063)} &\textcolor{black}{70.1 (0.014)}\\
2.0 &16.13 \textcolor{black}{(0.086)} &5.10 \textcolor{black}{(0.479)} &16.72 &23.77 &49.56 \textcolor{black}{(0.031)} &\textcolor{black}{206.6 (0.119)}\\
\end{tabular}
\end{ruledtabular}

\end{table*}

\begin{figure}

\includegraphics[scale=0.45]{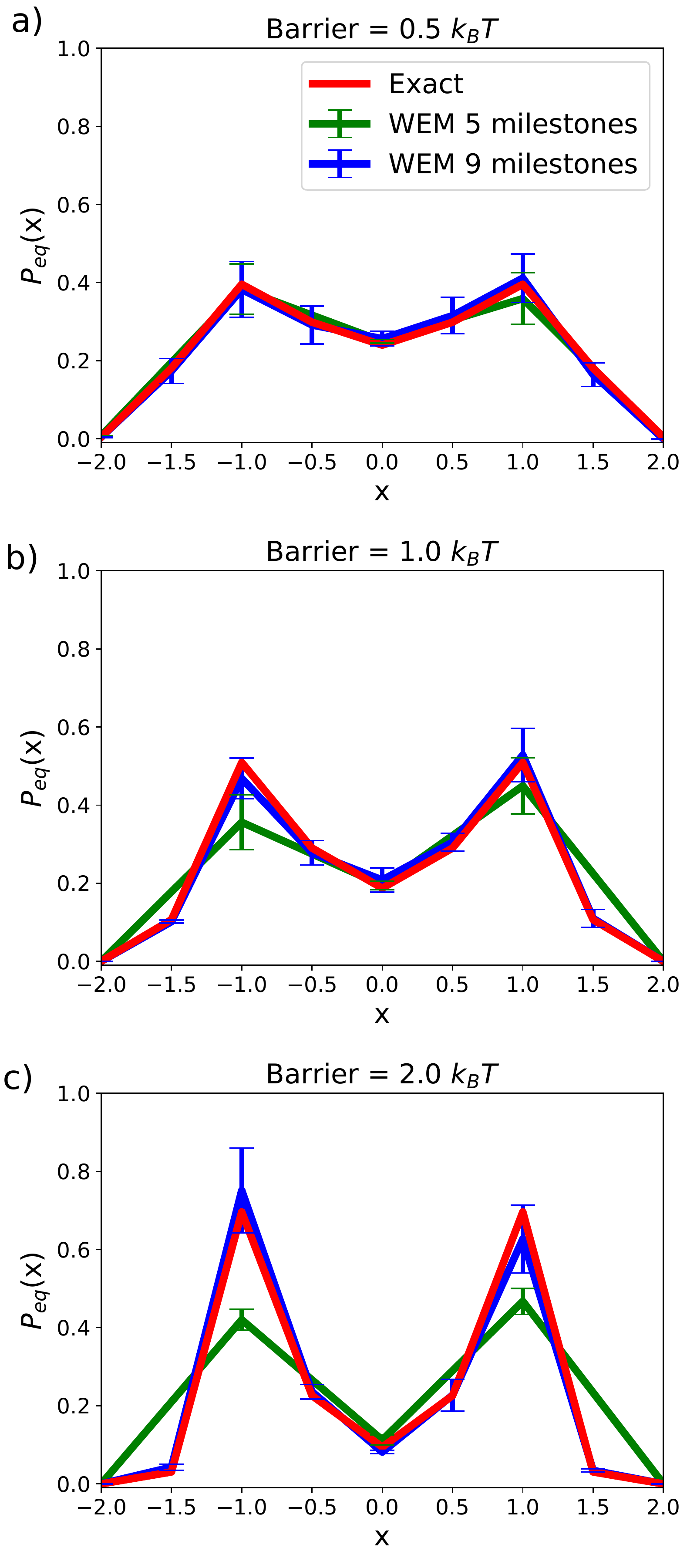}
\caption{Comparison of the stationary probability distribution of 1D double well systems, calculated from WEM simulations and exact Boltzmann probability. \textcolor{black}{Barrier heights are (a) 0.5 $k_BT$, (b) 1.0 $k_BT$ and (c) 2.0 $k_BT$.} 
}
\label{peq}
\end{figure}

The exact Boltzmann stationary distribution of probability along $x$ was compared with that computed from WEM simulation in Fig. \ref{peq}. The Boltzmann probability has been calculated only at the 9 milestone points ($x = -2.0 + 0.5i, i\in [0,8]$). 
The results of WEM simulation agree very well with the exact stationary distribution. The apparent discrepancy for the WEM 5 milestone case resulted due to the normalization of total probability over the milestones:
\begin{equation}
    \sum_{i=1}^M P_{\text{eq}}(x_i) \Delta x = 1
\end{equation}
If the total number of milestones changes, the normalized probability at a given milestone also changes. 

 Although better results can be expected with more number of milestones, very closely spaced interfaces will impose error in the calculation, since memory of previously visited milestones will not be completely lost. This trade off should be taken into consideration for biological application of this method. 

\begin{figure}
\includegraphics[scale=0.45]{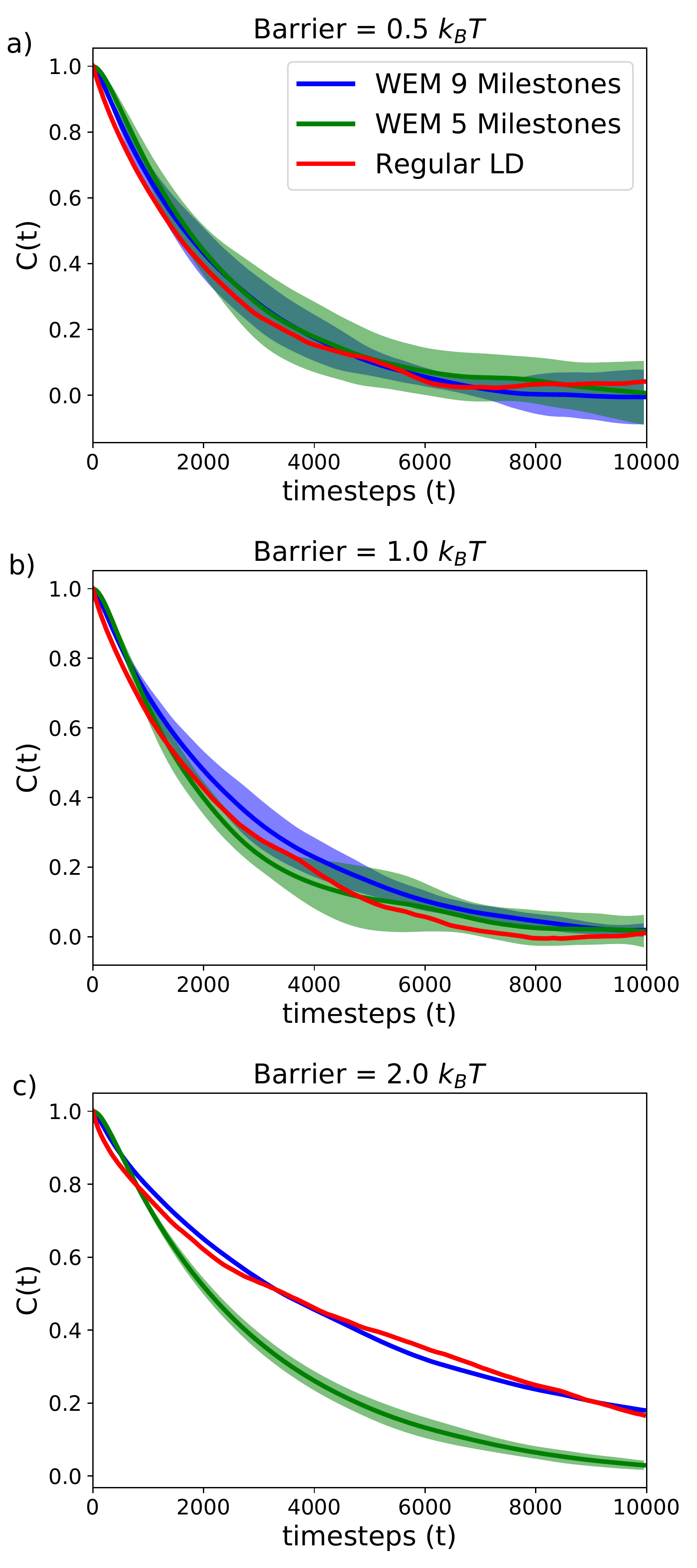}
\caption{Comparison of the position-position time correlation functions of 1D double well systems calculated from WEM simulations and regular Langevin dynamics (LD) simulation \textcolor{black}{for different barrier heights: (a) 0.5 $k_BT$, (b) 1.0 $k_BT$ and (c) 2.0 $k_BT$.}}
\label{time-corr}
\end{figure}

We implemented our interpolation-based method for calculating time correlation functions using Eq. \ref{eqn:time-corr} in Section \ref{sec:milestoning} for the short  milestone-to-milestone trajectories propagated using weighted ensemble scheme. We compared our results with a long brute-force Langevin dynamics simulation of $10^6$ steps for all different barrier heights. All trajectories were propagated long enough to obtain converged result for the time correlation function. Normalized position-position time correlation functions, $C(t) = \langle x(0)x(t) \rangle/\langle x(0)^2 \rangle$, calculated from WEM simulations with two different milestone configuration are shown in Fig. \ref{time-corr}. For lower barrier heights, e.g., $0.5k_BT$ and $1.0k_BT$, the $C(t)$ obtained from both milestone configurations closely resembles that from the long conventional Langevin dynamics simulation. For barrier height $2.0k_BT$ the $C(t)$ obtained from 5 milestones did not show quantitative agreement with Langevin dynamics result, but the 9-milestone case did. This is not surprising, as we have previously shown that the accuracy of $C(t)$ increases with increasing number of milestones \cite{Grazioli2018}. 

The results obtained for the 1D double well potential serves as a proof of concept for the  WEM method. We could show that it can reproduce the MFPT, stationary distribution and time correlation function with reasonable accuracy. The computational gain is not evident from our results for 1D potential because there is no additional degree of freedom. In atomistic molecular dynamics simulation there are many coupled degrees of freedom which the system will sample before transitioning to a new state along the reaction coordinate. In Section \ref{sec:11D} and \ref{sec:AA} we show that our method can be successfully applied in such scenarios as well. 

\subsection{(10+1)-dimensional Coupled Potential}
\label{sec:11D}
In order to study the effect of additional degrees of freedom coupled to the reaction coordinate, we have tested the WEM method on a (10+1)-dimensional potential \cite{Grazioli2018a} where the one dimensional reaction coordinate ($x$) is coupled with 10 low barrier double well potentials ($y_1,y_2, ...y_{10}$). The form of the potential function is
\begin{equation}
V(x,y_1,y_2,...y_{10}) = (1-x^2)^2 - \frac{1}{2} \sum_{n=1}^{10} y_n^2 x^2 + \sum_{n=1}^{10} y_n^4
\end{equation}
\begin{figure}
\includegraphics[scale=0.5]{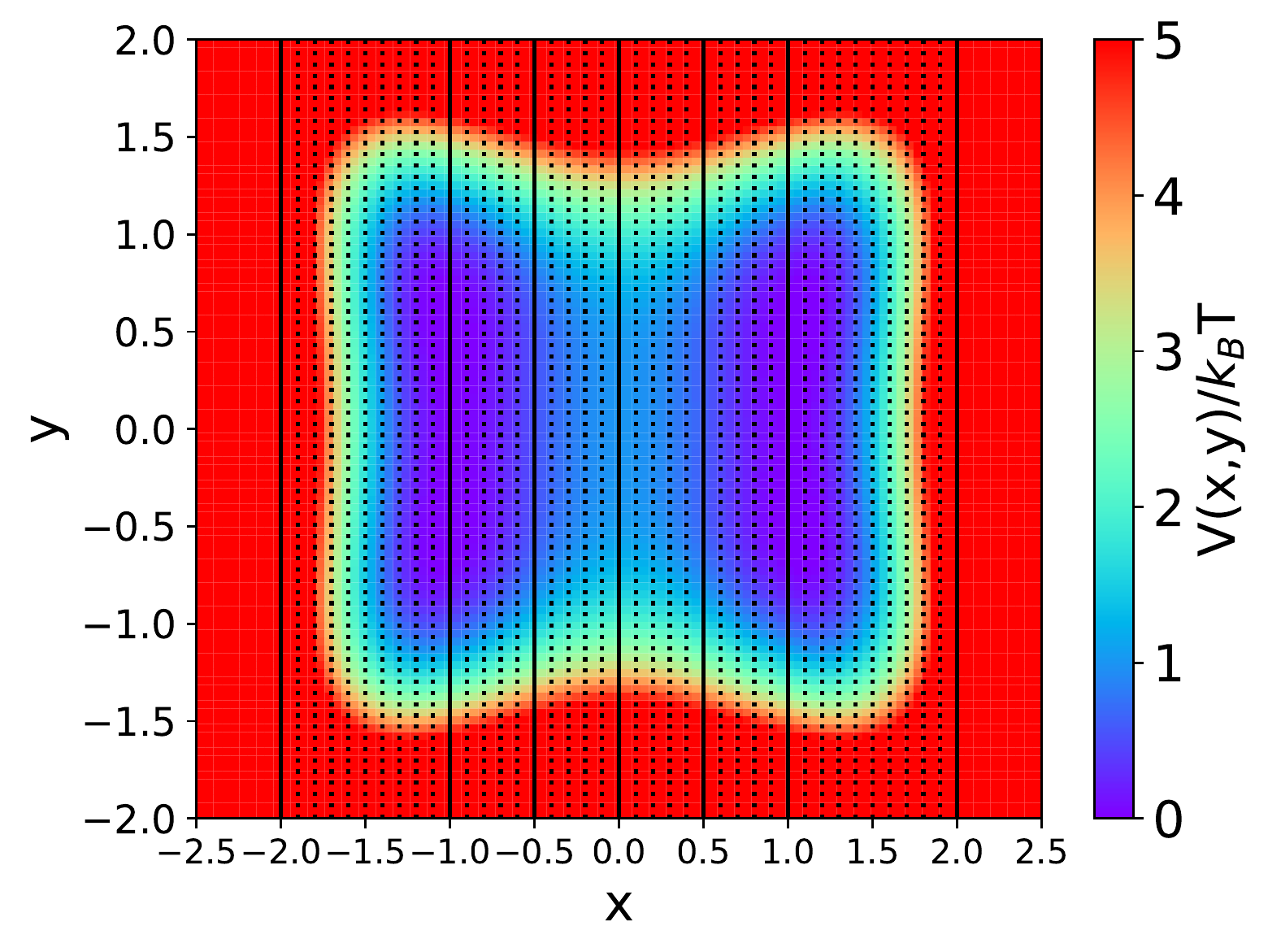}
\caption{A 2D projection of the (10+1)D coupled potential used in this study. In 2D the potential function is given by $V(x,y) = (1-x^2)^2 -0.5x^2y^2 + y^4 $. The black solid lines indicate milestones and the dotted lines indicate WE bin boundaries. The two additional milestones for 9 milestone case (at $x = \pm$1.5) are not shown in this figure. }
\label{11d}
\end{figure}
The projection of $V(x,y_1,y_2,...y_{10})$ on a 2D configuration plane is depicted in Fig. \ref{11d}. \textcolor{black}{Although symmetric in all coordinates, the potential along the orthogonal coordinates varies with $x$, resulting in a rugged energy landscape with $2^{11}$ minima. It works as a model for the complex energy surfaces characteristic of biomolecules. One ($x$) out of these 11 directions has a higher barrier and constitutes the slowest degree of freedom.}

Standard weighted ensemble (WE) simulations were performed without milestones with $\delta t = 500$ time steps for 1000 iterations. The WE bins were chosen to be of width $\delta x$ = 0.1 and trajectories were regenerated or stopped keeping 5 trajectories per occupied bin. Trajectories were initiated at $x = -1.0$ and stopped when they reached $x = 1.0$. For WEM simulations, milestones have been placed at $x =  -$2.0, $-$1.0, $-$0.5, 0.0, 0.5, 1.0, 2.0 for 7 milestone case and two additional milestones at $x = \pm$1.5 for the 9 milestone case. WE trajectories were propagated in between the milestones according to Equation \ref{eqn:odld}. Each WEM iteration involved $\delta t = 20$ time steps. All other parameters were same as those used for the 1D double-well potential. Three independent sets of simulation were performed for all WE and WEM simulations except for the 9 milestone case where we performed 6 independent sets of simulation as results showed larger variability.

\begin{figure}
\includegraphics[scale=0.5]{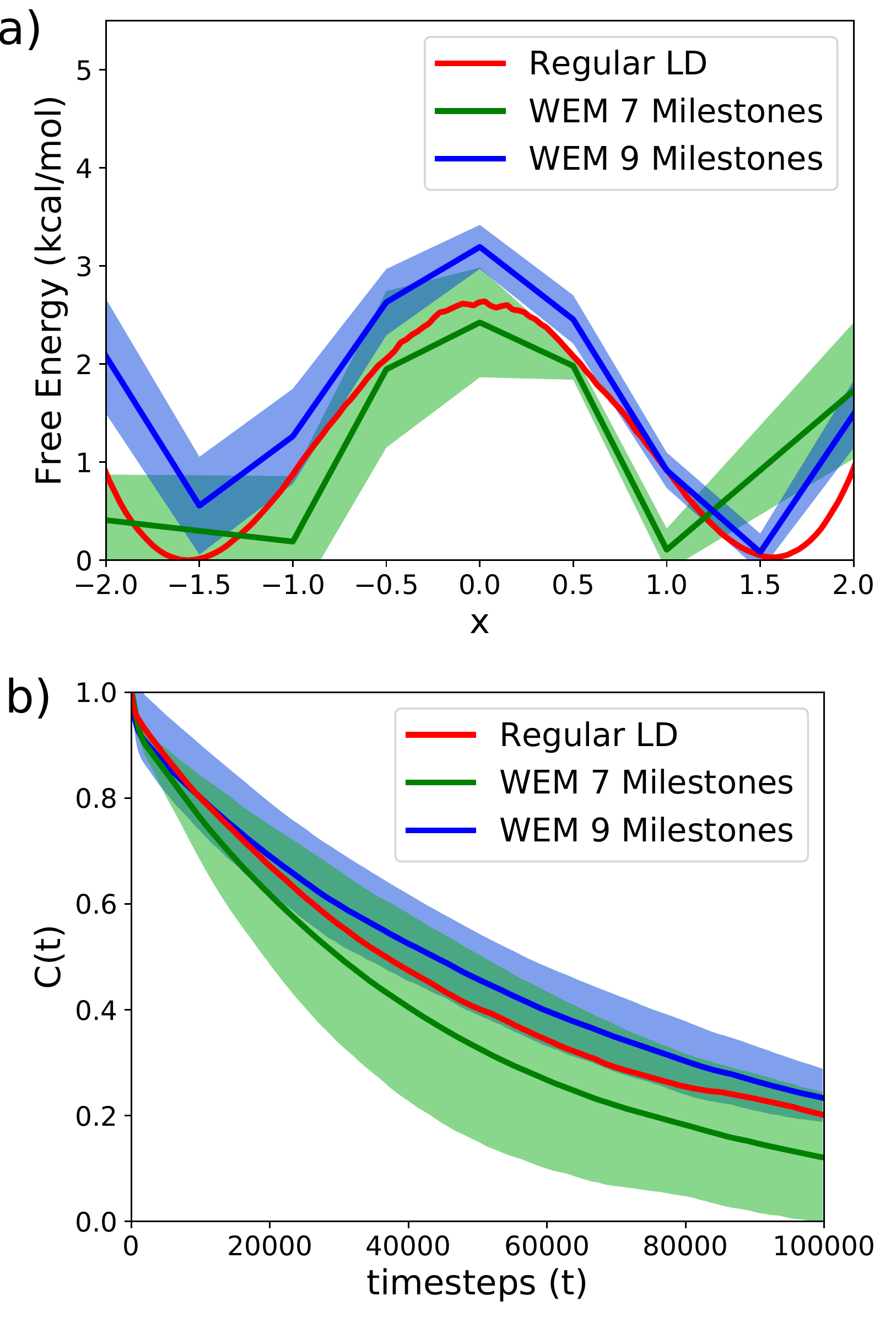}
\caption{(a) Free energy profile and (b) position-position auto-correlation function from the WEM scheme and from long regular Langevin dynamics (LD) simulation for (10+1) dimensional coupled potential. 
}
\label{11d-results}
\end{figure}

The transition matrix $\mathbf{K}$ and lifetime vector $\mathbf{\overline{T}}$ were computed from the WEM simulation to obtain the equilibrium probability distribution in milestone space. The free energy profile was computed from the probability distribution using Eq. (\ref{eqn:free-energy}). The results of both the seven- and nine-milestone schemes are well within 1 kcal/mol
agreement with that obtained from long regular Langevin dynamics simulation (Figure \ref{11d-results}). The results tend to improve with increasing number of milestones.

 The normalized position-position time-correlation function of the reaction coordinate $\langle x(0)x(t) \rangle/\langle x(0)^2 \rangle$ was computed from both WEM simulation and regular LD simulation and the results agree with each other (Figure \ref{11d-results}). The MFPT of transition from $x = -1$ to $x =1$ obtained from WEM simulation is also in agreement with that obtained from WE and regular Langevin dynamics (Table \ref{tab:11D}). \textcolor{black}{The algorithmic efficiency of WEM method is clearly better than regular LD or WE simulation. The convergence of each method is shown in the Supporting Information.} 
 
 These results indicate that our method can be extended to multi-dimensional systems with many degrees of freedom coupled to the reaction coordinate and experimental observables can be calculated in significantly less computational effort.

\begin{table}
\caption{\label{tab:11D} Mean first passage times (MFPT) for different simulation scheme for (10+1) dimensional coupled potential. Error bars are \textcolor{black}{95\% confidence interval} obtained from independent sets of simulation. Total simulation times required (including all walkers) to obtain converged MFPT results and \textcolor{black}{the inverse of algorithmic efficiency} have also been indicated.}
\begin{ruledtabular}
\begin{tabular}{ccccccc}
0Simulation scheme &MFPT &Total simulation time ($\eta^{-1}$)\\ &($\times 10^3$ time steps) &($\times 10^6$ time steps)\\
\hline
Regular LD &\textcolor{black}{105.2$\pm$29.7\footnote{Error bar is \textcolor{black}{95\% confidence interval} of 48 transition events observed in one long LD trajectory}} &10.0 \textcolor{black}{(0.797)}\\
WE &\textcolor{black}{93.5$\pm$16.6} &16.1\textcolor{black}{(0.507)}\\
WEM 7 milestone &\textcolor{black}{87.4$\pm$1.0} &1.9 \textcolor{black}{(0.0003)}\\
WEM 9 milestone &\textcolor{black}{120.1$\pm$54.8}
&1.1 \textcolor{black}{(0.229)}\\
\end{tabular}
\end{ruledtabular}

\end{table}

\subsection{Alanine Dipeptide}
\label{sec:AA}
To test the applicability of our method on biologically relevant systems we studied the transition between $\alpha_R$ and $C_{7eq}$ conformations in an artificially stiffened alanine dipeptide \cite{Frank2016,Nummela2007}. The 22-atom small molecule has been modeled with the CHARMM36 force field \cite{Huang2017} in a generalized Born implicit solvent (GBIS) environment \cite{AlexeyOnufriev2000,Onufriev2004}. Standard transition timescales and the free energy surface as a function of the backbone dihedral angles $\phi$ and $\psi$  have been computed from a long 1 $\mu s$ simulation. All MD simulations were performed using the NAMD 2.12 \cite{Phillips2005} package. Newtonian equations of motion were integrated with a 2 fs time-step with the SHAKE algorithm to constrain the bond lengths. We have artificially stiffened the backbone dihedral angle $\psi$ by applying harmonic walls to avoid the periodicity of the collective variable. Regular WE and WEM simulations were performed using the WESTPA \cite{Zwier2015} package. The $\psi$ space has been divided into bins of 10-degree width for WE simulation. The temperature has been kept constant at 300 K using a Langevin thermostat with a damping constant $\gamma =$ 80 ps$^{-1}$, which corresponds to water-like viscosity \cite{Adhikari2019ComputationalTimes}. The use of a stochastic thermostat is preferred over deterministic thermostats \cite{Zuckerman2017,Zuckerman2010}.
 
The WE simulation was performed for 500 iterations of $\delta t$ = 20 ps. The starting state was defined by $\psi = -60 ^{\circ}$ ($\alpha_R$) and the final state is defined at $\psi = 150 ^{\circ}$ ($C_{7eq}$). The $\phi$ dihedral angle was constrained between $-180^{\circ}$ and $0^{\circ}$ by a harmonic wall constraint of 0.04 kcal/mol deg$^{-2}$. Three independents sets of WE simulation were performed and the mean and 95\% confidence interval of all observables were reported.

For WEM simulations 8 milestones were placed at the following $\psi$ angles: -100$^{\circ}$, -60$^{\circ}$, -20$^{\circ}$, 20$^{\circ}$, 60$^{\circ}$, 100$^{\circ}$, 150$^{\circ}$ and 180$^{\circ}$. For each of the milestone the starting point of the simulation is generated by performing a 100 ps equilibration simulation starting from the energy minimized structure. During equilibration the $\psi$ angle was constrained at the specific value of the milestone by a force constant of 0.12 kcal/mol deg $^{-2}$. The same bins and $\gamma$ were used for the WE simulation. WE trajectories were initiated from the final structure of the 100-ps simulation and propagated until its daughter trajectories reached either of the nearby milestone. An iteration time $\delta t$ of 0.2 ps was used. Mean and 95\% confidence interval of MFPT, free energy and time-correlation function were calculated from six independent sets of simulation.

\begin{figure}
\includegraphics[scale=0.5]{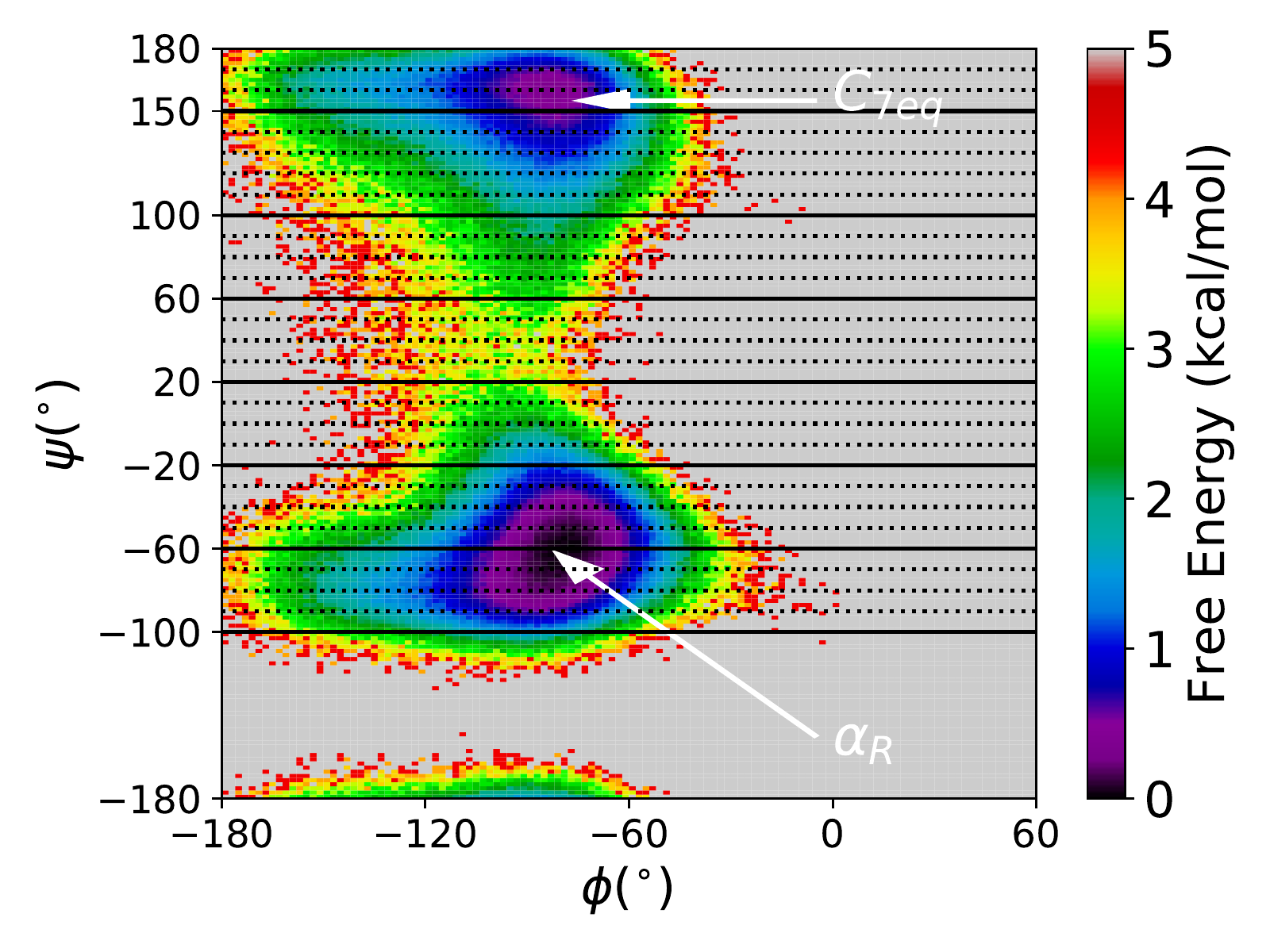}
\caption{Free energy profile of the artificially stiffened Alanine dipeptide obtained from 1 $\mu$s long conventional MD simulation. The position of the milestones are depicted by solid horizontal lines and the edges of the WE bins are shown by dotted lines.}
\label{AA-freeE}
\end{figure}
The mean first passage times for transition between $\alpha_R$ and $C_{7eq}$ states have been summarized in Table \ref{tab:AA} for all three methods. The total simulation times required to obtain converged results have also been shown (Table \ref{tab:AA}). The MFPT and its standard deviation (not shown in table) from conventional MD simulation show very similar values, indicating a Poisson distribution of timescales for barrier-crossing events. This Poisson type behavior agrees with previous work by Salvalaglio et al. \cite{Salvalaglio2014} As mentioned in Section \ref{sec:1d}, the computational gain is clearly pronounced for a molecular system with many degrees of freedom. The total simulation time for all walkers for WEM simulation is $\sim$30 times less than a conventional WE calculation. Moreover, the WEM method can be parallelised over milestones. The longest time required to get converged transition probability and lifetime on a single milestone is about 3-4 ns. So, provided the availability of the parallel computing resources, one effectively spends two orders of magnitude less wall clock time for WEM simulation than a traditional WE or MD simulation. The mean first passage time for an $\alpha_R$ to $C_{7eq}$ transition is in reasonable agreement with the results from long MD and WE simulations (Table \ref{tab:AA}). \textcolor{black}{The algorithmic efficiency defined in Eq. (\ref{eqn:efficiency}) is also much better for the WEM method when compared to conventional MD and WE simulation. The convergence of first passage times using each method is shown in the Supporting Information.} 

In traditional WE implementations, we need to specify a starting and a target state. This prevents us from obtaining the rates of the backward process without performing a second set of calculations. \textcolor{black}{(There are exceptions to this as described in Section \ref{sec:concluding-discussion})}. As an additional advantage over the traditional WE method, the mean first passage time of the backward transition (here, $C_{7eq} \rightarrow \alpha_R$) can also be estimated by using the transpose of transition kernel $\mathbf{K}^T$ and $\mathbf{\overline{T}}^{\prime}$ in place of $\mathbf{K}$, and $\mathbf{\overline{T}}$, respectively, in Eq.  (\ref{eqn:mfpt}), where 
\begin{equation}
    \overline{T}^{\prime}_i = \overline{T}_{M-i}; \;\;\;\;\;\; i \in [ 1,M ]
\end{equation}
with $M$ the number of milestones. The predicted timescales of the $C_{7eq} \rightarrow \alpha_R$ transitions are in order-of-magnitude agreement with the result obtained from a 1-$\mu$s long, regular MD simulation.
\begin{table}
\caption{\label{tab:AA} Mean first passage times (MFPT) and computational costs for obtaining converged results from different simulation schemes for alanine dipeptide. 
}
\begin{ruledtabular}
\begin{tabular}{ccccccc}
Simulation  &MFPT (ns) &MFPT(ns) &Total simulation \\
scheme &($\alpha_R \rightarrow C_{7eq}$) &($C_{7eq} \rightarrow \alpha_R$) &time \textcolor{black}{($\eta^{-1}$)} (ns)\\
\hline
Brute Force MD\footnote{Error bar is \textcolor{black}{95\% confidence interval of 277} transition events observed from a long trajectory} &\textcolor{black}{2.4$\pm$0.3} &\textcolor{black}{1.2$\pm$0.1} &1000 \textcolor{black}{(15.6)}\\
WE &\textcolor{black}{2.1$\pm$0.5} &- &545 \textcolor{black}{(30.9)}\\
WEM &\textcolor{black}{3.0$\pm$1.6}
&\textcolor{black}{0.7$\pm$0.5} &11.6 \textcolor{black}{(3.3)}\\
\textcolor{black}{WEM \footnote{Error analysis performed using the method described in appendix. Only reported for one of the trials. The absolute MFPT of the particular trial is reported followed by the 95\% confidence interval in paranthesis}} &\textcolor{black}{2.0(0.1)} &\textcolor{black}{0.9(0.02)} &-
\end{tabular}
\end{ruledtabular}

\end{table}

Moreover, about 4.5 ns simulation time was spent in the last two milestones ($\psi = 150.0^{\circ}$ and $180.0^{\circ}$). If calculating the rate constant is the only objective, these calculations are not necessary, because the transition probabilities from the target milestone and beyond do not appear in the modified transition kernel $\mathbf{\tilde{K}}$ in Eq. (\ref{eqn:ktilde}). But for the calculation of the free energy profile and time correlation functions, transition probabilities and lifetimes at these milestones are required. Also, 100 ps of additional equilibration was performed to sample initial equilibrium configurations at each milestone. These added up to 800 ps more simulation time to that reported for WEM simulations in Table \ref{tab:AA}.

\begin{figure}
\includegraphics[scale=0.47]{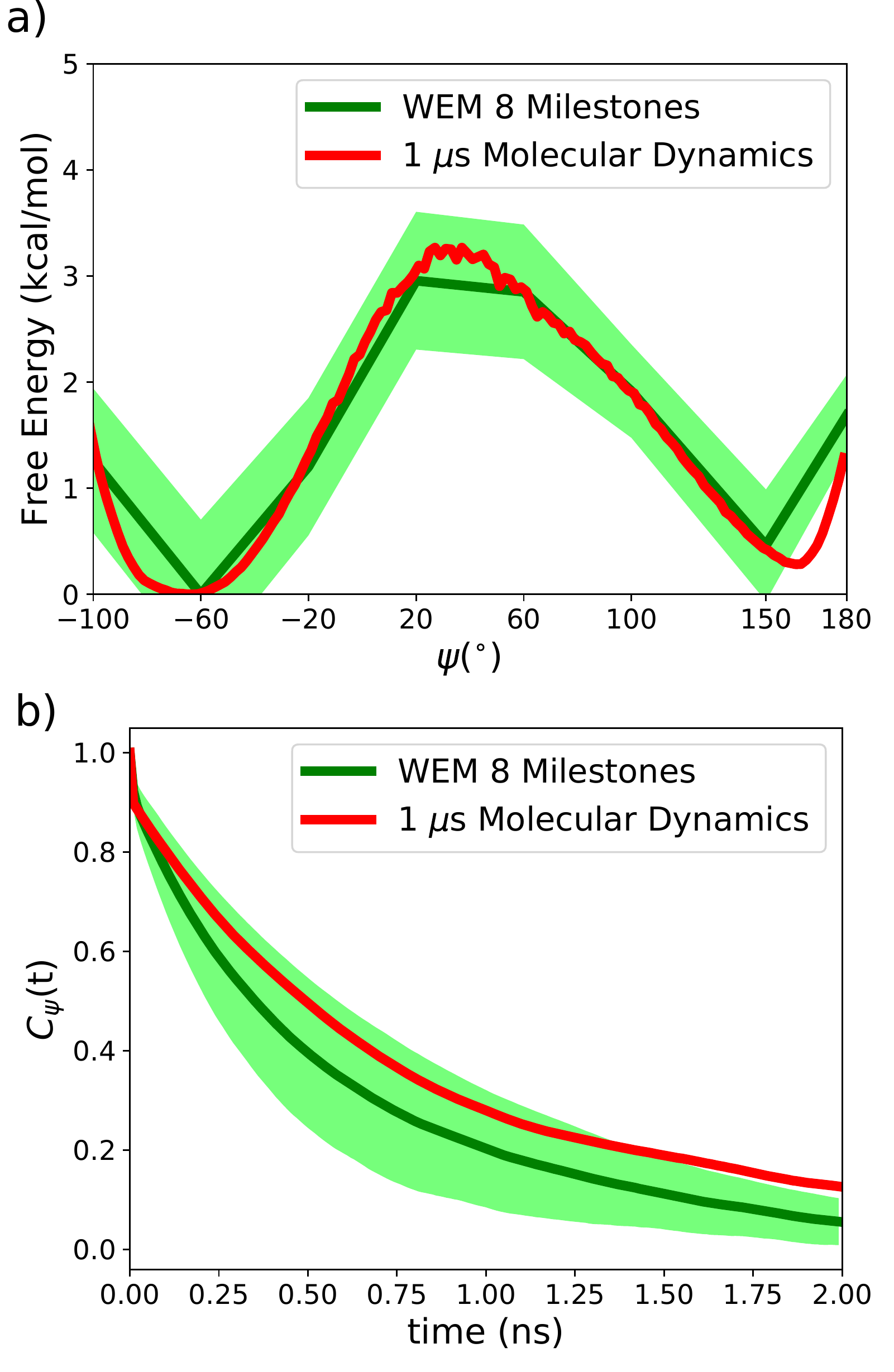}
\caption{Comparison of the (a) free energy profile along $\psi$ and (b) the time correlation function   $C_{\psi}(t)$ calculated from WEM simulation with 8 milestones and long  conventional MD simulation. The error bars are \textcolor{black}{95\% confidence interval} computed from 6 independent WEM simulations.}
\label{AA_results}
\end{figure}

\begin{figure}
\includegraphics[scale=0.5]{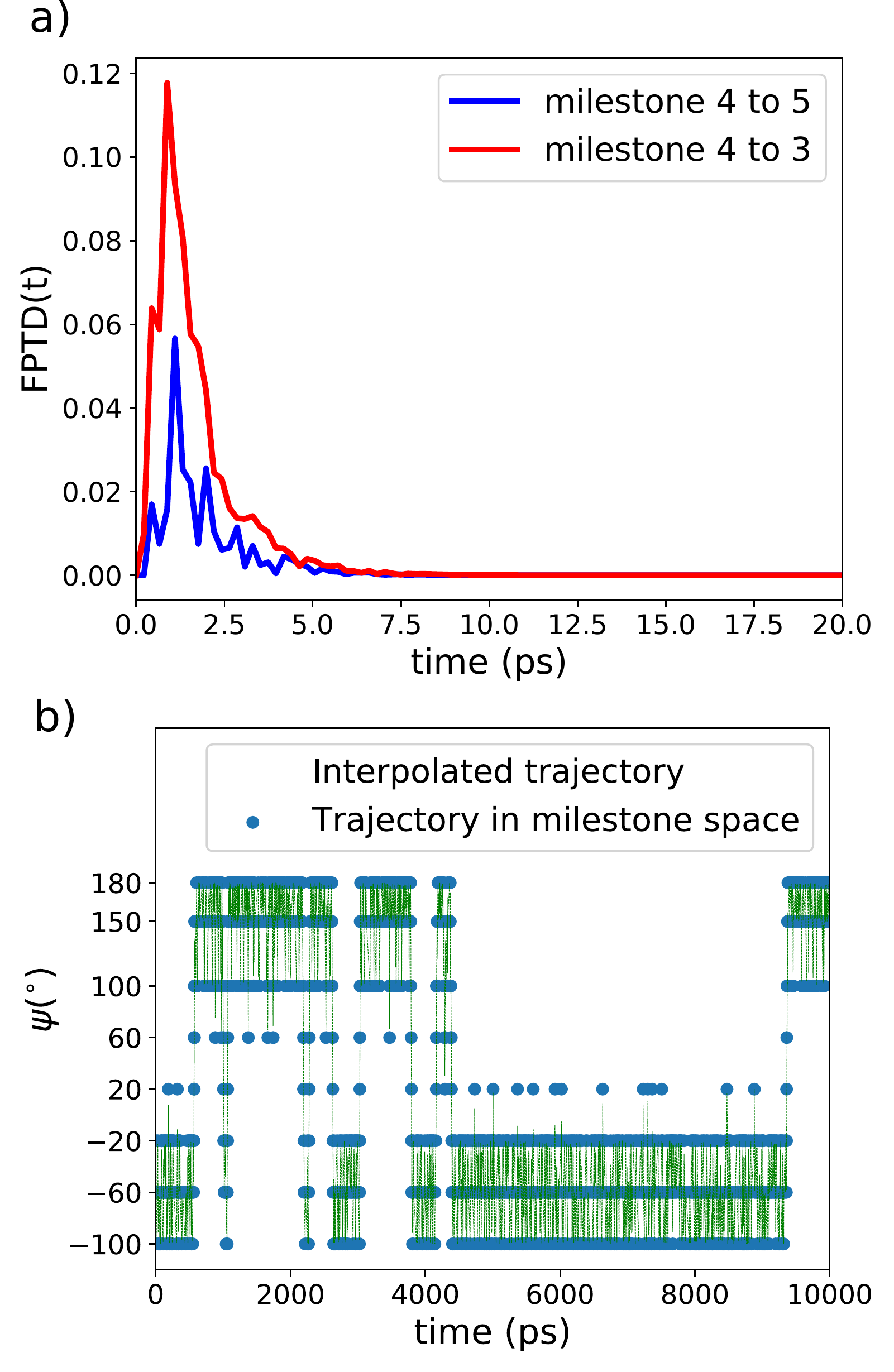}
\caption{(a) Convergence of the First Passage Time Distribution function from milestone 4 ($\psi = 20^{\circ}$) to milestone 3 ($\psi = -20^{\circ}$) and milestone 5 ($\psi = 60^{\circ}$). (b) The milestone to milestone trajectory and the interpolated continuous trajectory generated from the first passage time distribution between milestones (as described in Section \ref{sec:milestoning})}.
\label{time_corr_AA}
\end{figure}

The stationary probability distribution in milestone space was calculated and the values were used in Eq. (\ref{eqn:free-energy}) to obtain the free energy profile. The reference ($P_{\text{eq}}^0$) was chosen to be the probability of the most populated milestone. The free energy profile from WEM is in quantitative agreement with the one computed from long unbiased MD simulation (Figure \ref{AA_results}). \textcolor{black}{West \textit{et al.} have shown that the free energy profile obtained from conventional milestoning of molecular systems agrees very well with that obtained from long equilibrium MD simulation \cite{West2007}. Our results indicate that this agreement is extended to the WEM method as well.} We could also predict the barrier height to be 2.9 kcal/mol, which is very close to the value obtained from a conventional MD simulation (3.3 kcal/mol). Also, the normalized auto-correlation function of the $\psi$ dihedral angle $\left(C_{\psi}(t) = \frac{\langle \psi(0)\psi(t) \rangle}{\langle \psi (0)^2 \rangle}\right)$ was computed as a function of time and and compared with the results of the 1-$\mu$s MD simulation. The results agree very well with each other (Figure \ref{AA_results}).  The convergence of the first passage time distribution ($FPTD(t)$) for one of the milestones is depicted in Fig. \ref{time_corr_AA}. The milestone-space trajectory generated from the transition statistics, and the interpolated trajectory used for the calculation on the time correlation function are shown in Fig. \ref{time_corr_AA}.

\section{Concluding Discussion}
\label{sec:concluding-discussion}
In this paper we have developed and tested a combined weighted ensemble milestoning (WEM) method for calculating kinetic and thermodynamic properties for systems with long-tims barrier crossing events. We have tested our model for a 1D double well model with different barrier heights, a 11D potential with 10 degrees of freedom coupled to reaction coordinate (RC), and an atomistic model of Alanine dipetide. We have computed stationary probability distribution, free energy profile along RC,  mean first passage time and time correlation function for these systems. The WEM simulations were able to reproduce the results of regular MD or Langevin dynamics, conventional milestoning, and weighted ensemble simulations within a small fraction of the computing time. Also, the possibility of parallelizing the calculation over different milestones reduces the wall clock time by two orders of magnitude compared to regular MD or WE simulation. Moreover, WEM also allows the calculation of the mean first passage time and rate for the \emph{reverse} process, which is not directly available from conventional WE calculations \textcolor{black}{performed under steady-state conditions (with trajectory regeneration from the initial state) or equilibrium conditions. However, it has been shown that transition timescales between arbitrary states can be computed from a different implementation of the WE scheme \cite{Suarez2014SimultaneousTrajectories}}.

WEM has superficial similarities with the adaptive weighted ensemble procedure (aWEP) by Bhatt and Bahar \cite{Bhatt2012}, who also aim to enhance WE simulations. However, the two are fundamentally different methods. Unlike the aWEP method, WEM utilizes the background of milestoning to calculate the free energy profile and underlying time correlation functions. Also, the primary aim of WEM is not to obtain equilibrium rate constant in between milestones, but to calculate the transition kernel $\mathbf{K}$, the lifetime vector $\mathbf{\overline{T}}$, and the first passage time distribution between each pair of milestones.

Unlike the WARM method we proposed \cite{Grazioli2018a}, WEM does not require application of biasing forces, although, in principle,the strategy could be additionally introduced. The unidirectional wind forces used in WARM can increase sampling in the forward direction, but, on the other hand, decrease sampling on the backward direction. For example, applying a biasing force along $+x$ direction for our 1D model will cause more trajectories to reach milestone $i+1$ from milestone $i$ with very less number of trajectories reaching milestone $i-1$, resulting in significant reduction of the statistics for $i\rightarrow i-1$ transition. For calculating observable properties like free energy and kinetics from milestoning,
 it is necessary to properly sample both back and forth transitions. Our WEM method does not create any directional bias and increases sampling in all directions.  
 
We have previously also developed a technique for calculating time correlation function by decomposing a single long trajectory into milestones \cite{Grazioli2018}. Here, we have extended this technique for short milestone to milestone trajectories and showed the applicability of their method for enhanced milestoning simulations. Pure WE simulation can not reproduce time correlation functions very well because of the correlated nature of the trajectories as the daughter trajectories from the same parent have the exact same evolution history \cite{Zuckerman2017}. We showed that it is possible to recover the correct time correlation functions from WE simulation using milestoning. 

A typical problem for milestoning simulation is the optimal placement of the milestones. If they are placed too close to each other, the transition times might be shorter than the relaxation timescales. This causes preservation of the memory of previously visited milestones and the master equation based formalism becomes no longer applicable.
To avoid this, spacing between milestones need to be increased so that their transition statistics are independent. This, however, leads to significant increase in the length of the trajectories and computational cost. WEM decreases the computational cost of simulating transitions between such distant milestones by performing weighted ensemble simulations instead of traditional molecular dynamics.

In the conventional milestoning scheme, one needs to perform an equilibrium simulation on each milestone constraining the reaction coordinate and allowing sampling of the other orthogonal degrees of freedom. Then, a large number of trajectories (usually of the order of $10^2 - 10^3$) are to be generated from different starting points sampled from that equilibrium trajectory. But WEM does not necessarily require one to start many trajectories from each milestone. As one WE trajectory splits into many trajectories with smaller weights, one or a few starting trajectories are sufficient to reasonably sample the space in-between milestones and also the transition probabilities. So starting points can be obtained from a very short constrained equilibration at each milestone. All our calculations were performed with just one starting point from each milestone, but a few more might be required for problems with more complex energy landscapes, like the ones involved in protein folding or other significant rearrangements at each milestone. However, the number of starting trajectories required is significantly less compared to regular milestoning, bringing enormous simplicity in the simulation workflow.

\textcolor{black}{Given a relatively small number of starting points, the initial structures on each milestone may not properly reflect the equilibrium distribution of the coordinates orthogonal to the progress variable. We realized that this may be the primary source of error in our calculation. Therefore, we performed multiple independent sets of calculation for all systems to quantify the variability rather than to estimate error bars from Monte-Carlo bootstrapping (results only shown for alanine dipeptide).}

It is here of interest to compare with related findings concerning one-long vs. many short trajectories as a general sampling strategy. Earlier studies confirmed that the process of decomposing a long equilibrium trajectory into many faster, but non-equilibrium ones accelerates the convergence of free energies \cite{Sun2003EquilibriumTrajectories,Ozer2010AdaptiveY}. In another example, in early work on MD applications for proteins, it was found that a single 5 ns long trajectory of crambin could only explore a fraction of the conformational space sampled from ten independent 120 ps trajectories with same initial coordinate, but different starting velocities . \cite{Caves1998LocallyCrambin} Similar considerations are reflected in MSM \cite{Pande2010} and WExplore simulations \cite{Dickson2014WExplore:Algorithm}, which show that running multiple trajectories starting from different or similar initial condition significantly increases the amount of conformational space explored. One needs of course to be below the ergodicity limit for such differences to exist, but in essence, it this spawning of multiple new starting points that is the origin of enhancement in our stratification-within-stratification WEM method in comparison to continuous dynamics.

Lastly, we note that as it is, our method does not directly identify the choice of optimal reaction coordinates for the entire process, which in itself is an extremely involved exercise in biomolecular systems. However, once a good order parameter space is known, it can help in reaching correct physics by spending very small amount of computational effort. Moreover, once the milestones are laid down, self-averaging techniques \cite{Andricioaei2003Self-guidedAverages,MacFadyen2005ATrajectories} can be employed to enhance the weighted ensemble part along the slow manifold \textit{in-between} milestones. 

In summary, we proposed \textcolor{black}{a novel rare-event sampling strategy, named weighted ensemble milestoning (WEM), which combines the efficiency of WE with the theoretical framework of milestoning to estimate kinetics, free energy profiles and time correlation functions from short discontinuous trajectories. In this paper we provide a proof-of-principle study for our method and we test it on a few model systems.}
We have implemented our method in WESTPA toolkit \cite{Zwier2015WESTPA:Analysis} through NAMD molecular simulation package \cite{Phillips2005}. The method can be extended to higher dimensional milestones through Voronoi bins as suggested by Vanden-Eijnden and coworkers \cite{Vanden-Eijnden2009}. We hope that WEM will find application in computational biophysics in deciphering thermodynamics and kinetics of complex and challenging processes, and that it will help characterize rare event physics in biomolecular systems in the future.

\section*{Supplementary material}

\textcolor{black}{See supplementary material for the convergence plots of LD, MD, conventional WE and WEM simulations.}

\section*{Acknowledgements}
The authors thank Trevor Gokey for his help in generating the scripts for analyzing the WEM results. We are grateful to Moises Romero, Shane Flynn, Bridgett Kohno, Brian Ngyuen, and Saswata Roy for their valuable comments and suggestions. IA acknowledges the generous support of National Science Foundation via grant CMMI 1404818. The work has benefited from the computational resources of the UC Irvine High Performance Computing (HPC) cluster.

\section{Data availability statement}
The software codes of the WEM implementation are available from the authors upon request. The data that supports the findings of this study are available within the article. Additional data files can be made available from the authors upon request.

\appendix
\section{Error analysis}
\textcolor{black}{There are different methods of error analysis for weighted ensemble and milestoning based methods. As all the quantitative results we report are obtained using the milestoning equations, we restrict our discussion to the error analysis of the milestoning based techniques.} 

\textcolor{black}{Given a milestoning trajectory data set there are two commonly used methods to estimate error. The simplest one is the error propagation scheme in which the standard deviation of the mean first passage time and free energy values are obtained from the stationary flux and standard deviation of the lifetime of individual milestones \cite{Tang2020TransientDesign}. The key formulae for this method are from the supporting information of Ref. \citen{Tang2020TransientDesign}. But obtaining confidence intervals using this method is difficult as the sample size is not well defined.}

\textcolor{black}{A more rigorous alternative is estimating statistical errors by generating a distribution of transition matrices using the Bayesian-type conditional probability proposed in Refs. \citen{Majek2010} and \citen{Vanden-Eijnden2008} and elaborately described in Ref. \citen{Votapka2015}. The probability of obtaining a rate matrix $\mathbf{Q}$ given the transition counts and lifetime is given by 
\begin{equation}
    p(\mathbf{Q}|\lbrace N_{ij},\overline{T}_{i} \rbrace ) = \prod_i \prod_{j \neq i} Q_{ij}^{N_{ij}} \exp{(-Q_{ij} N_i \overline{T}_i)} P(\mathbf{Q})
    \label{eqn:likelihood}
\end{equation}}

\textcolor{black}{Here $N_{ij}$ is the number of trajectories starting from milestone $i$ and hitting milestone $j$ before hitting any other milestone, and $N_i$ is the total number of trajectories staring from milestone $i$. The $P(\mathbf{Q})$ is an uniform prior. The rate matrix $\mathbf{Q}$ is defined such that the transition kernel $\mathbf{K}$ and lifetime vector $\mathbf{\overline{T}}$ can be obtained from it using
\begin{equation}
    \begin{split}
        &K_{ij} = \frac{Q_{ij}}{\sum_{l \in \lbrace i-1,i+1 \rbrace} Q_{il}  } \;\;\;\; (i \ne j) \\
        &\overline{T}_i = \frac{1}{\sum_{l \in \lbrace i-1,i+1 \rbrace} Q_{il}}
    \end{split}
    \label{eqn:QKT}
\end{equation}
}

\textcolor{black}{In conventional milestoning, the elements of $\mathbf{Q}$ are obtained by maximizing the likelihood described in Eq. (\ref{eqn:likelihood}). 
\begin{equation}
    Q_{ij} = \frac{N_{ij}}{N_i \overline{T}_i} \;\;\;\; (i \ne j)
\end{equation}
This formula is not directly applicable to WEM calculation because the weights of the trajectories are different. By analogy with Equation \ref{eqn:kij} we define the elements of $\mathbf{Q}$ for WEM to be 
\begin{equation}
    Q_{ij} = \frac{K_{ij}}{\overline{T}_i} \;\;\;\; (i \ne j)
    \label{eqn:qij_wem}
\end{equation}
where $\mathbf{K}$ is computed using Eq. (\ref{eqn:kij_wem}). It is clear that Eq. (\ref{eqn:qij_wem}) satisfies Eq. (\ref{eqn:QKT}). The diagonal elements are defines as $Q_{ii} = - \sum_{i \ne j} Q_{ij}$.} 

\textcolor{black}{Now the posterior in Equation (\ref{eqn:likelihood}) is sampled by a non-reversible element shift Monte Carlo algorithm \cite{Noe2008ProbabilityModels} following Ref. \citen{Votapka2015} to generate multiple samples of the rate matrix labelled as $\mathbf{Q}'$. In this algorithm, one of the off diagonal non-zero elements of $\mathbf{Q}$ is chosen randomly and a modification $\Delta$ to the element is made in the following way
\begin{equation}
    \begin{split}
        &Q_{ij}' = Q_{ij} + \Delta \;\;\;\; (i \ne j)\\
        &Q_{ii}' = Q_{ii} - \Delta.
    \end{split}
    \label{eqn:q_move}
\end{equation}
The second line of Equation \ref{eqn:q_move} is necessary to preserve the row stochasticity of the matrix. The displacement $\Delta$ is randomly drawn from an exponential distribution of the range $\Delta \in [-Q_{ij},\infty)$ with a mean of zero, to ensure that the elements of the rate matrix preserve their signs. The $\mathbf{Q}'$ matrix obtained this way is accepted with a probability
\begin{equation}
    p_{accept} = \frac{p(\mathbf{Q}|\lbrace N_{ij},\overline{T}_{i} \rbrace )}{p(\mathbf{Q'}|\lbrace N_{ij},\overline{T}_{i} \rbrace )} = \left( \frac{Q_{ij} + \Delta}{Q_{ij}}\right)^{N_{ij}} \exp{(-\Delta N_{i} \overline{T}_i )}
\end{equation}
Each of the accepted rate matrices are converted back to $\mathbf{K}$ and the mean first passage times and free energies are computed. The 95\% confidence interval obtained using the $z$-statistic is reported as the error bar.}
\section*{References}
\bibliography{references}

\end{document}